\documentclass[aps,onecolumn,nofootinbib,superscriptaddress,letterpaper,amsmath,amssymb]{revtex4}
\pdfoutput=1

\usepackage{silence}
\usepackage{amsmath}
\usepackage{graphicx}
\usepackage{dcolumn}
\usepackage{bm}
\usepackage{amssymb}
\usepackage{feynmf}
\usepackage{slashed}
\usepackage{multirow}
\usepackage{color}
\usepackage{array}
\usepackage{url}
\usepackage{tikz}
\usepackage{siunitx}
\usepackage{mathtools}
\usepackage{subcaption}
\usepackage{float}
\usepackage{makecell}
\usepackage{lmodern}
\usepackage{anyfontsize}
\usepackage{comment}

\usepackage[utf8]{inputenc} 
\usepackage[T1]{fontenc}    
\usepackage{url}            
\usepackage{booktabs}       
\usepackage{amsfonts}       
\usepackage{nicefrac}       
\usepackage{microtype}      
\usepackage{xcolor}         
\usepackage{hyperref}       

\begin{document}

\title{Fast multi-geometry calorimeter simulation with conditional self-attention variational autoencoders}

\author{Dylan Smith}
\email{dylanrs@uci.edu}
 \affiliation{Department of Physics and Astronomy, University of California, Irvine, CA 92697}
\author{Aishik Ghosh}
 \affiliation{Department of Physics and Astronomy, University of California, Irvine, CA 92697}
 \affiliation{Physics Division, Lawrence Berkeley National Laboratory, Berkeley, CA 94720}
\author{Junze Liu}
 \affiliation{Department of Computer Science, University of California, Irvine, CA 92697}
\author{Pierre Baldi}
 \affiliation{Department of Computer Science, University of California, Irvine, CA 92697}
\author{Daniel Whiteson}
 \affiliation{Department of Physics and Astronomy, University of California, Irvine, CA 92697}


\begin{abstract} 

The simulation of detector response is a vital aspect of data analysis in particle physics, but current Monte Carlo methods are computationally expensive. Machine learning methods, which learn a mapping from incident particle to detector response, are much faster but require a model for every detector element with unique geometry. Complex geometries may require many models, each with their own training samples and hyperparameter tuning tasks. A promising approach is the use of geometry-aware models, which condition the response on the geometry, but current efforts typically require cumbersome full geometry specification. We present a geometry-aware model that takes advantage of the regularity of detector segments, requiring only the definition of cell sizes across regular segments. This model outperforms the current state of the art by over 70\% across several key metrics including the Wasserstein distance metric. 
\end{abstract}

\maketitle

\section{Introduction} 
\label{intro}

Simulating expected detector response is a key component of data analysis in particle physics experiments~\cite{Cranmer:2019eaq}. Simulated events are vital to assess the expected power of experiments and plan upgrades to current detectors as well as predict the response of hypothetical future detectors.

Monte Carlo (MC)-based simulation, such as Geant4~\cite{GEANT4:2002zbu,Allison:2016lfl,Allison:2006ve},  models the initial interaction of a particle with the detector as well as the secondary and tertiary byproducts. This approach can provide very a high-fidelity description, but comes with significant computational cost.
 Simulation of the calorimeter is especially expensive, as incident particles create a cascading shower of particles which must be modeled. As a result, the computational cost for simulation is a scientific bottleneck for experiments with complex detectors. Additionally, increasing data volume and complexity from future and upgraded  detectors will  exacerbate this issue.

Machine-learning (ML) based simulation tools directly learn a mapping between incident particles and detector responses, which can be substantially faster than MC-based approaches~\cite{baldi2014searching,baldi2021deep,calafiura2022artificial,lu2022resolving,shmakov2023end,Paganini:2017dwg, 3dgan_epj,Buhmann:2020pmy,Krause:2021ilc,Krause:2021wez,Mikuni:2022xry, amram2023denoisingdiffusionmodelsgeometry}. However, the response depends sensitively on the detector geometry, such that irregular geometries may require many specialized models~\cite{ATLAS:2021pzo,ATLAS:2022jhk,Erdmann:2018jxd,Ratnikov:2020dcm,Hashemi:2023ruu}, and rapid prototyping of future detectors is hindered by the need to generate MC-based training samples for each hypothetical geometry. Geometry-aware approaches~\cite{liu2022geometryaware}  mitigate this issue by learning how the detector response depends on the geometry, allowing simulation of a geometry out of the training data. An important obstacle is that these approaches typically require full specification of the geometry, including the position and size of every calorimeter cell. However, most detectors have significant regions of regularity, allowing potential reduction in ML model complexity. In this paper, we present a geometry-aware calorimeter simulation in which the geometry is defined in regular sections, avoiding full geometry specification. The model successfully interpolates between training geometries, which allows the for accurate simulation of detector images for geometries it has not been trained on, for which MC-based simulation is not required.  

This paper is organized as follows. Sec.~\ref{bkgrd} provides the context for ML-based simulation. Sec.~\ref{dataset} describes the samples of simulated detector response used in the paper. Sec.~\ref{arch} discusses the model architecture and training. Sec.~\ref{results} presents the performance of the model relative to the state of the art, and Sec.~\ref{conclusion} summarizes the work.

\section{Background} 
\label{bkgrd}

Simulation of the detector response to particles produced in interactions is central to data analysis in particle physics, including inference about parameters of the Standard Model (SM) or setting limits on theories of new physics. The calorimeter, which measures the energy of incident particles, is a crucial element of the detector, but the complexity of the cascading showers requires enormous computing resources to simulate with standard Monte Carlo tools such as Geant4. However, such tools provide high-fidelity simulations and naturally handle the complex and irregular geometry of the calorimeter, which can feature many different cell sizes and shapes.

Machine learning approaches instead learn to map the incident particle directly to a detector response, without a detailed model of the microphysics that generates the response. This can be much faster, but as the response depends sensitively on the detector geometry, it can require as many networks as there are unique geometries. For example, AtlFast3~\cite{ATLAS:2021pzo} trains a distinct network for each detector component. This approach requires significant human effort to train and validate each fixed-geometry network, which does not immediately transfer to any new task, such as modeling the response of upgraded detectors or hypothetical new detectors~\cite{dg2024}.

Machine learning has demonstrated strong performance in parameterized learning~\cite{Baldi_2016},  which learns the dependence of the task on some parameter and is able to perform well even for data outside of its training set. Geometry-aware simulation capitalizes on this capacity by learning detector response as a function of cell geometry. In principle, this would allow it to generate expected detector responses for new geometries, without requiring generation of expensive samples and validation of new networks. As such, a single such geometry-aware model could replace hundreds of individual fixed geometry models. 

One approach to geometry-aware simulation avoids dependence on the geometry altogether by modeling detector deposition as continuous points~\cite{Kansal:2021cqp,Mikuni:2023dvk,Leigh:2023toe}, which can later be grouped into arbitrary cells. In practice, these point-cloud models learn extremely granulated uniform detector segmentation rather than truly continuous responses. This high segmentation presents a more complex learning task with significant computational cost.  An alternative approach is an autoregressive model (ARM) such as Geometry-Aware Autoregressive Model (GAAM)~\cite{liu2022geometryaware}, which iterates over cells sequentially, predicting each cell based on previous cells, allowing for conditioning on the geometry of each cell.

\begin{figure}
    \centering
    \includegraphics[width=0.75\textwidth, trim=0 5cm 0 5cm]{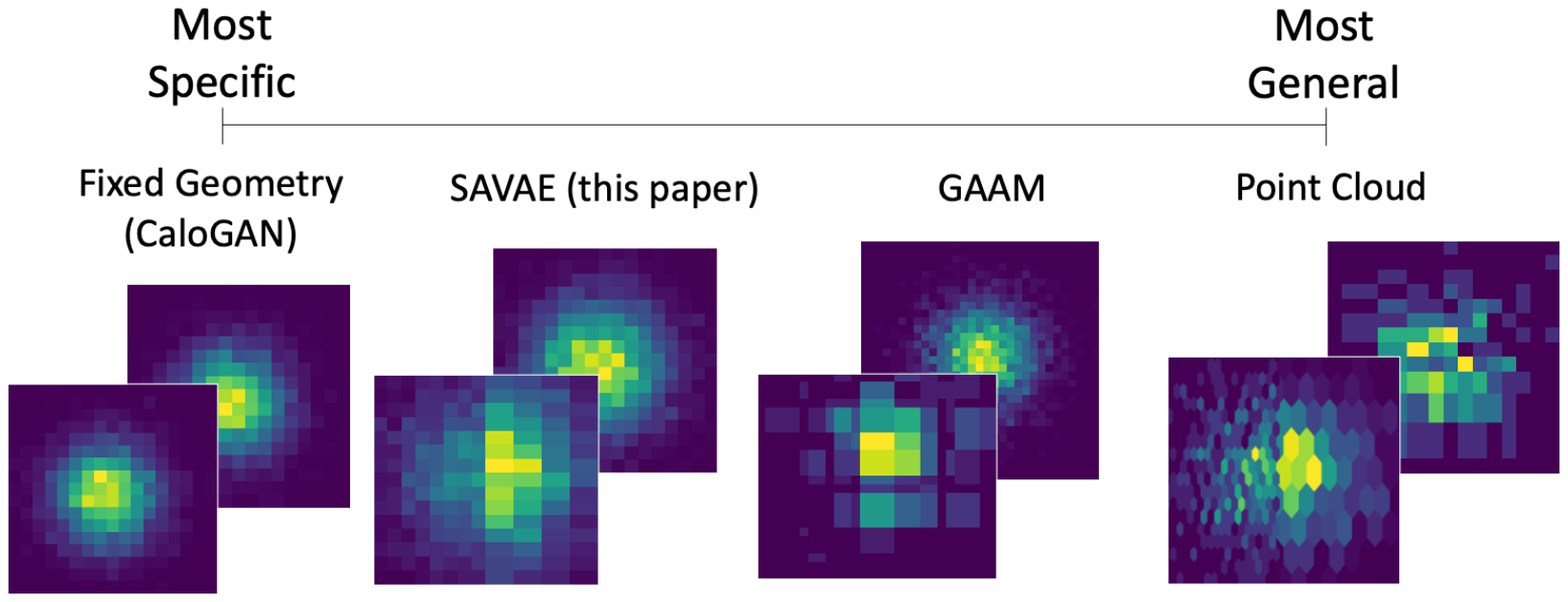}
    \caption{ Comparison of calorimeter-generating ML models by their degree of geometry-specificity.  Fixed geometry models such as CaloGAN assume train for specific geometries. The geometry-aware SAVAE model learns the response for any regions of regular rectangular geometry. GAAM learns to simulate any rectangular cell shape. Point Cloud methods learn to simulate any geometry, even hexagonal. }
    \label{fig:spectrum}
\end{figure}

Both point-cloud and autogressive models are powerful and fully general, capable of modeling even a calorimeter in which each cell is unique, but their generality  may be more than the task calls for, as both approaches come with increased computational costs. In reality, many calorimeters have large regions of regularity, and models that capitalize on this behavior can reduce complexity and cost, minimizing the number of models trained without paying the full price of the most general geometry-aware models.  This modelling philosophy can be thought of as a spectrum (shown in Fig.~\ref{fig:spectrum}) for which a balance can be achieved in terms of the generality of the model and its performance.

\section{Dataset}
\label{dataset}

The dataset used is the same used to train GAAM~\cite{liu2022geometryaware}. Samples of simulated calorimeter response used for training and evaluation are generated with very fine segmentation, approximating the infinite segmentation of point-clouds, to allow for later grouping into cells of various sizes and shapes without resimulation. Segmentation follows the strategy of the CaloGAN dataset~\cite{Paganini_2018}, but with 16-fold more segmentation.

The simulated calorimeter is composed of three longitudinal layers labeled `inner', `middle', and `outer' layers along the $z$-axis. The transverse coordinates are $\eta$ and $\phi$, and specific geometries are indicated by $(n_{\eta},\ n_{\phi})$, where $n_{\eta}$ is the number of cells along the $\eta$-axis and $n_{\phi}$ is the number of cells along the $\phi$-axis. To study the effectiveness of the geometry-aware model on several detector geometries, the point-cloud images are grouped into cells of varying length and width, described in Table~\ref{tab:geo_table}. Each layer has an physical area of $48 \times 48$ cm$^2$, and the layers have a length of 5mm, 40mm, and 80mm in $\eta$, 160mm, 40mm and 40mm in $\phi$ and a depth of 90mm, 347mm and 43mm respectively. Note that the depth of each layer is disregarded in the model; a single 3D calorimeter shower can be decomposed as three 2D images, one for each calorimeter layer.

\begin{figure}
     \centering
     \begin{subfigure}[b]{0.3\textwidth}
         \centering
         \includegraphics[width=\textwidth]{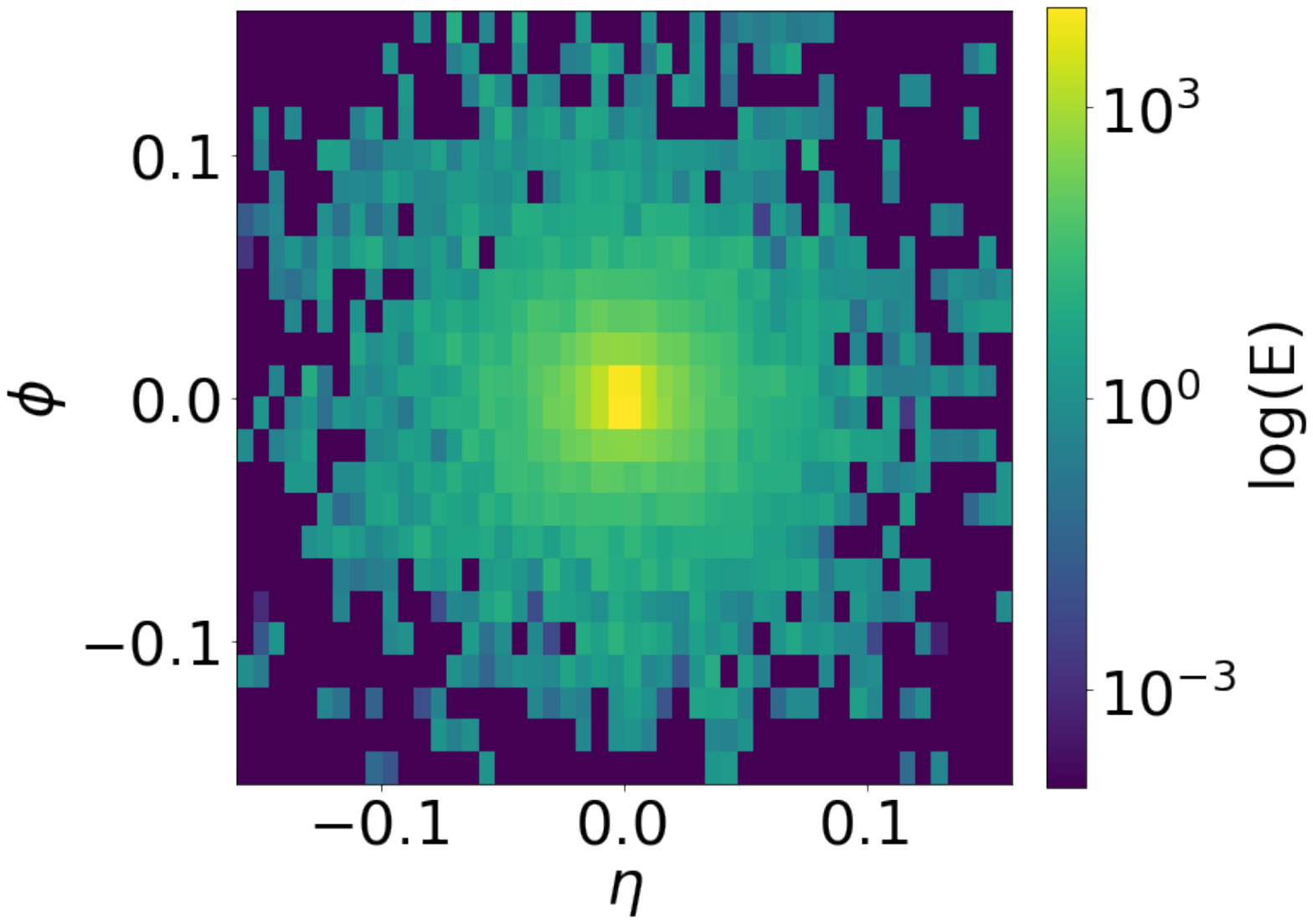}
         \caption{(48,24)}
         \label{fig:48x24_same}
     \end{subfigure}
     \hfill
     \begin{subfigure}[b]{0.3\textwidth}
         \centering
         \includegraphics[width=\textwidth]{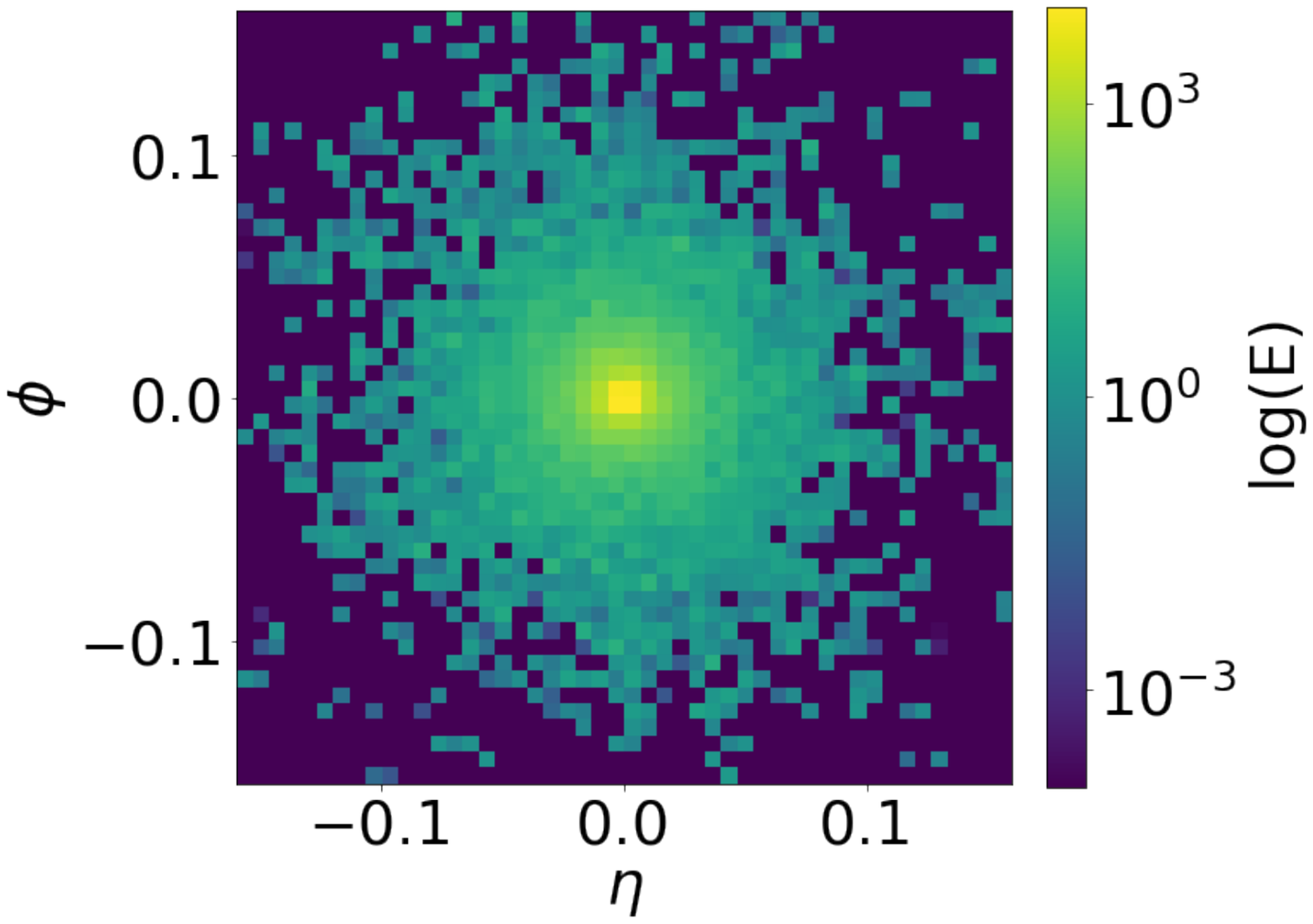}
         \caption{(48,48)}
         \label{fig:48x48_same}
     \end{subfigure}
     \hfill
     \begin{subfigure}[b]{0.3\textwidth}
         \centering
         \includegraphics[width=\textwidth]{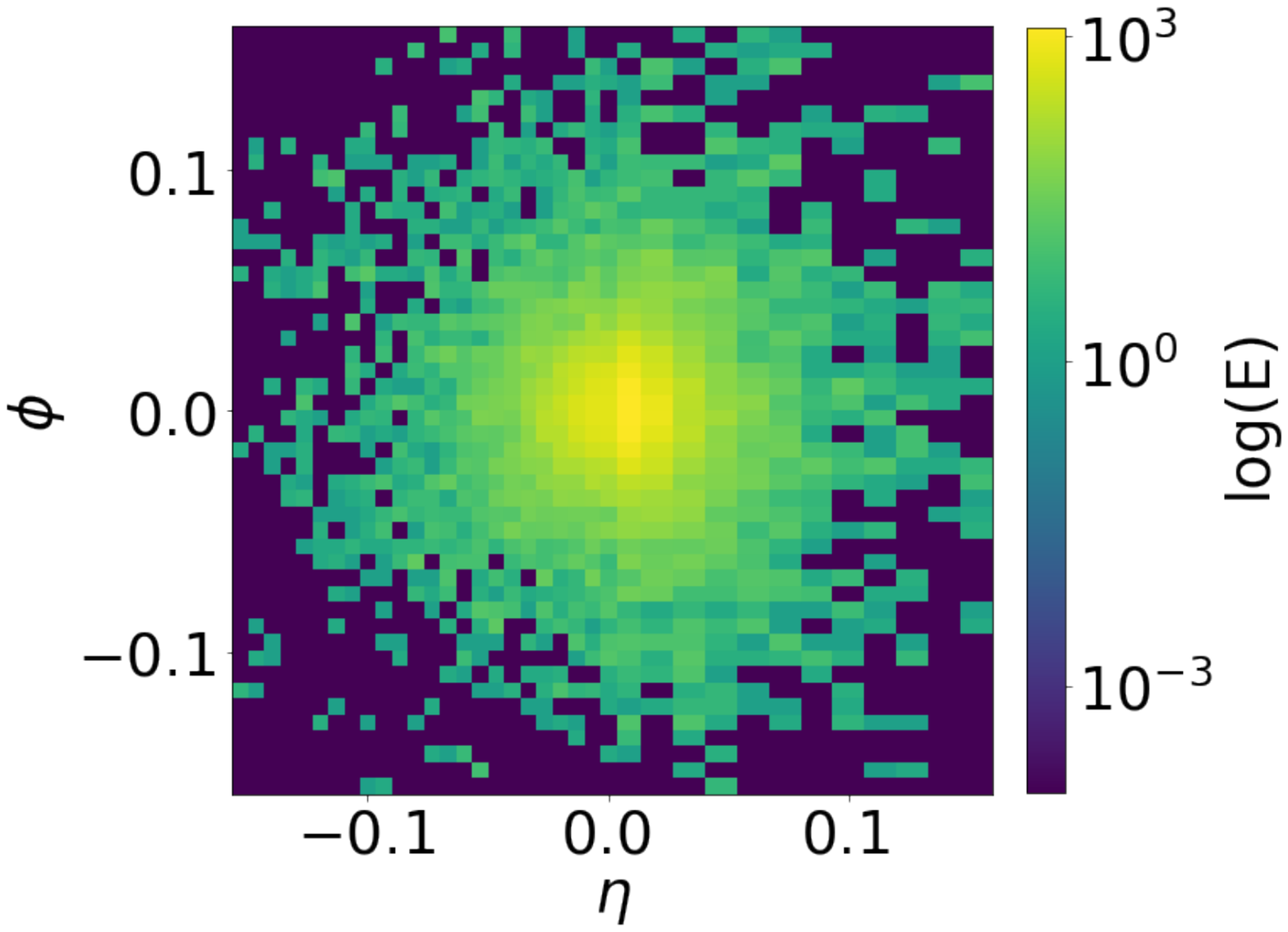}
         \caption{(36,48)}
         \label{fig:36x48_same}
     \end{subfigure}
     \caption{Preparation of various geometry configurations from a single high-granularity shower generated by Geant4. The nonuniform geometry of Fig.~\ref{fig:36x48_same} represents the transition region of the detector.}
     \label{fig:sameshower}
\end{figure}

This grouping is demonstrated visually in Fig.~\ref{fig:sameshower}, in which a single layer calorimeter image is grouped into different geometries. Note that in the $(36,48)$ geometry, the right half of the image is segmented into a (12, 48) grid and the left half into (24, 48) cells in $(n_{\eta}, n_{\phi})$, and so cell size is not uniform over the entire image. This geometry represents a transition region where the detector geometry changes abruptly.

A sample of 10,000 photons with fixed energy of 65 GeV are generated and used for training. The beam is focused on the center of detector $(0,0)$ in $\eta/\phi$-space. As the photon hits the calorimeter, it creates a shower of secondary photons and electrons that deposit energy onto the calorimeter layers. The middle layer receives the largest amount of energy, while the outer layer receives the least and is incredibly sparse as a result.

\begin{table}
    \centering
    \caption{Geometry of each layer of the simulated calorimeters used in training the model and evaluating the model's ability to interpolate unseen geometries, indicated by $(n_{\eta},\ n_{\phi})$, the number of cells in $\eta$ and $\phi$, respectively. Several configurations are used, including one (marked with a *) in which the cell division is not uniform.}
    \label{tab:geo_table}
    \begin{tabular}{cc}
          \toprule
          \bf{Image Type} & \bf{Calorimeter Segmentation} \\ 
          \midrule
          Training & (12,12), (48, 24), (48,48), (36,48)*\\ 
          \midrule
          Unseen & (24,24), (24,12) \\  
          \bottomrule
    \end{tabular}
\end{table}

For evaluation on unseen geometries, additional Geant4 images are binned into geometries the model was not trained on to assess the model's ability to interpolate to unseen geometries. These interpolated geometries are $(24,24)$ and $(24,12)$, and are generated for the middle layer of the calorimeter.

\section{Model}
\label{arch}

A wide variety of generative model architectures have demonstrated the capacity to simulate the calorimeter response for specific geometries, such as graph neural networks (GNNs)~\cite{4700287}, generative adversarial networks (GANs)~\cite{Paganini_2018} and variational auto-encoders (VAEs)~\cite{kingma2022autoencoding}.  For application to geometry-aware learning, models must also be able to condition on the cell size and handle training data with varying cell multiplicity.

Several architectures were explored without success. A Graph Neural Network Variational Autoencoder, in which calorimeter image cells are represented as nodes of a graph connected only to neighboring nodes, performed well for fixed geometries but failed to learn the dependence of the response across geometries. This is likely a result of local connections failing to capture global features.  A GAN approach suffered from well-known training instability and mode collapse~\cite{salimans2016improved} for fixed geometries, which were exacerbated for the more general problem and thus would require a much greater degree of optimizing the model hyperparameters.  

The basic structure of a VAE includes an encoder and a decoder neural network~\cite{kingma2022autoencoding}. The encoder maps input data to a lower-dimensional latent space representing the mean and variance of a Gaussian distribution. The decoder maps from the latent space back to the data space. The model is trained to minimize a reconstruction error $R$, which compares the input data to the data after the encoder and decoder mappings, as well as a Kullback-Leibler (KL) divergence $D_{KL}$ which measures the Gaussian nature of the latent space. These two terms are balanced by a hyper parameter $\beta$~\cite{higgins2017betavae} such that the total loss is  $L=R+\beta D_{KL}$.  Generating a new image requires only generating a sample from a Gaussian and mapping it with the decoder.    VAEs are typically simpler to train and to generate new samples than GANs or ARMs~\cite{lu2021sparse}.  However, some adaptation of the VAE architecture is required to handle highly-granular images, to  learn the dependence on cell geometry by capturing long-range relationships, and to ensure diversity in the generated samples.

\begin{figure}
     \centering
     \begin{subfigure}[t]{0.4\textwidth}
         \centering
         \includegraphics[width=\textwidth]{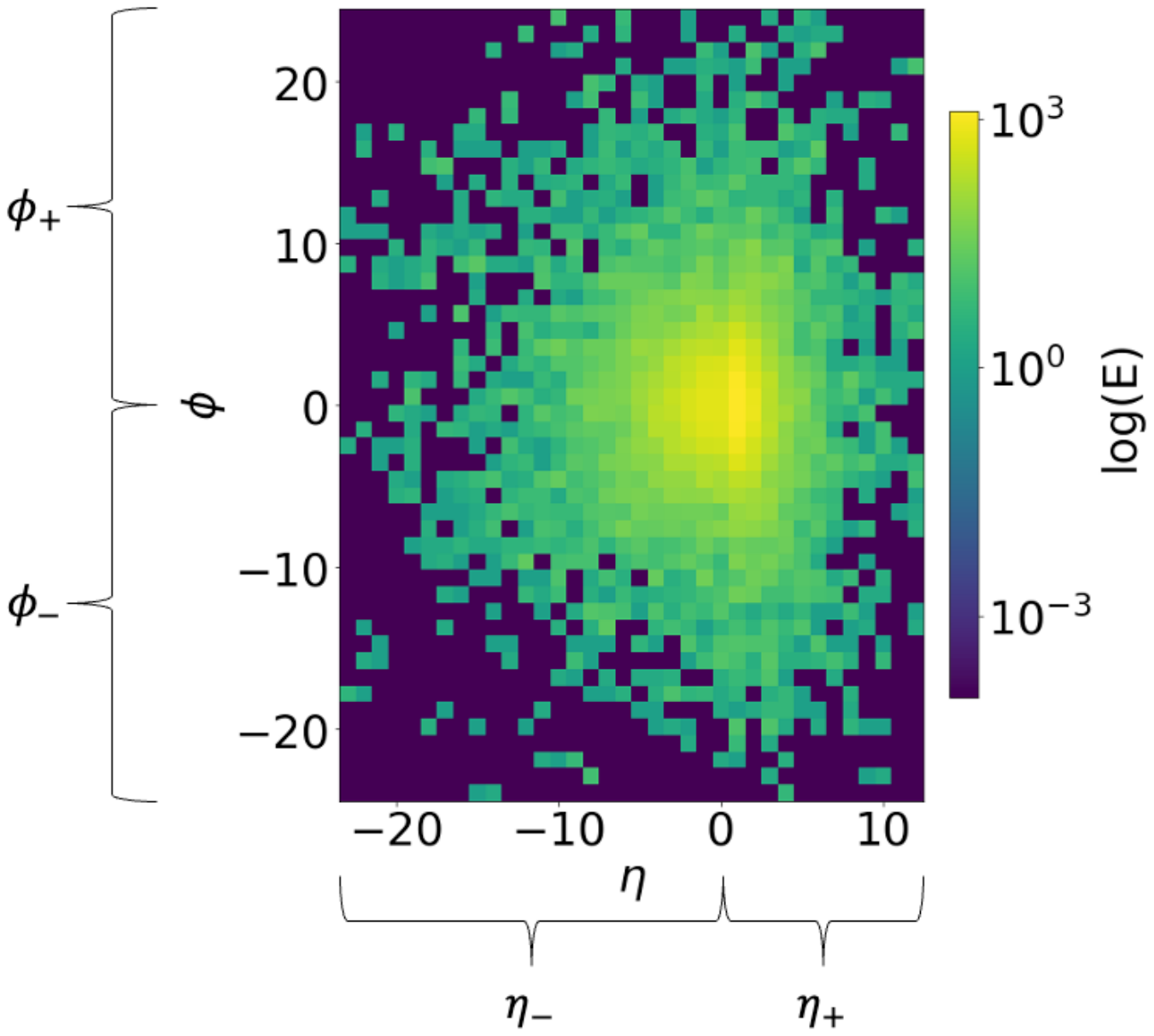}
         \caption{Geometry Conditioning}
         \label{fig:cond}
     \end{subfigure}
     \hfill
     \begin{subfigure}[t]{0.5\textwidth}
         \centering
         \includegraphics[width=\textwidth]{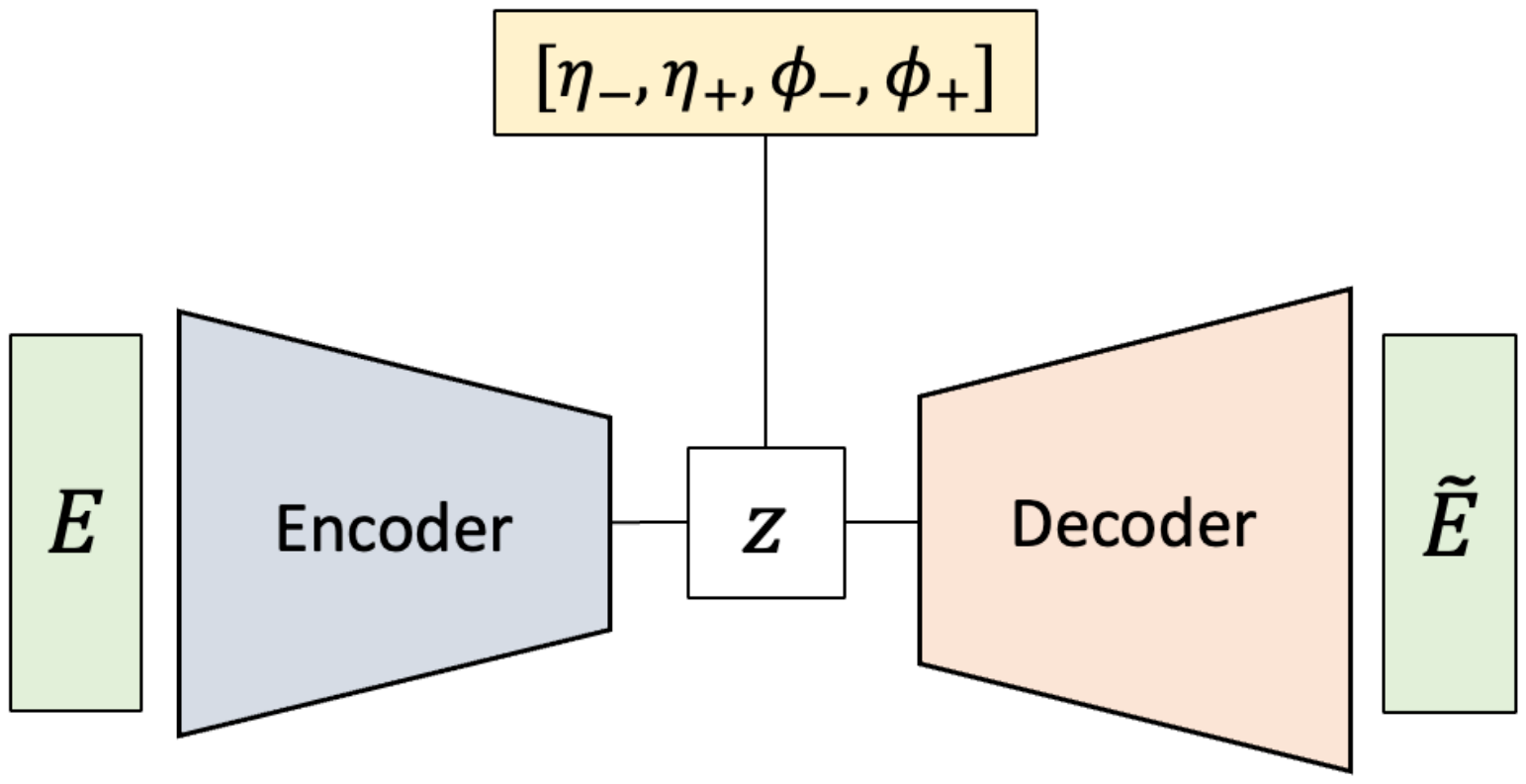}
         \caption{SAVAE}
         \label{fig:arch}
     \end{subfigure}
     \caption{(a) Demonstration of specification of the geometry via a conditioning vector defined in the text. Note that the cells are scaled to be the same size to better visualize the number of cells for $\eta<0$ and $\eta>0$. (b) SAVAE architecture diagram, where $E$ is the array of cell energies, $\tilde{E}$ are the reconstructed cell energies, and $\bf{z}$ is the latent embedding}
     \label{fig:model_setup}
\end{figure}

To learn energy deposition as a function of cell size, the model is conditioned on the calorimeter geometry, represented by the vector $[\eta_-, \eta_+, \phi_-, \phi_+]$ and concatenated to the latent embedding. The value   $\eta_-$ ($\eta_+$) gives the number of cells for $\eta<0$ ($\eta>0$), and similarly for $\phi_-$ ($\phi_+$). For example, a $(48,48)$ geometry with regular cells would have a conditioning array $[24,24,24,24]$. For the non-uniform geometry $(36,48)$, the  conditioning vector is $[24,12,24,24]$ (shown in Fig.~\ref{fig:cond}). An example and the model architecture is shown in Fig.~\ref{fig:arch}. 

Self-attention, which allows neural networks to weigh the importance of cells  by considering their relationships to each other~\cite{zhang2019selfattention}, is implemented to facilitate learning both local and global features. Such models learn long-range dependencies more effectively compared to convolution architectures, even in variable-length samples. The encoder of the self-attention VAE (SAVAE) is a single self-attention layer, followed by two dense layers with ReLU activation. The decoder is two dense layers with ReLU activation and one dense layer with sigmoid activation.

The simulated calorimeter energy deposits are preprocessed by taking the logarithm and scaling cell energies to the range $[0,1]$. Training was performed on a single A500 NVIDIA GPU for 500 epochs. The hyperparameters are the latent space dimensionality, the number of training epochs, the KL factor $\beta$, and the number of self-attention layers. Training was especially sensitive to the $\beta$-term, with an optimal value found of $10^{-4}$. Larger values of $\beta$ led to decreased diversity in image samples, and failure to interpolate to unseen geometries.

\section{Results}
\label{results}

\begin{figure}
    \centering
    \begin{subfigure}[b]{0.45\textwidth}
        \centering
        \includegraphics[width=\textwidth]{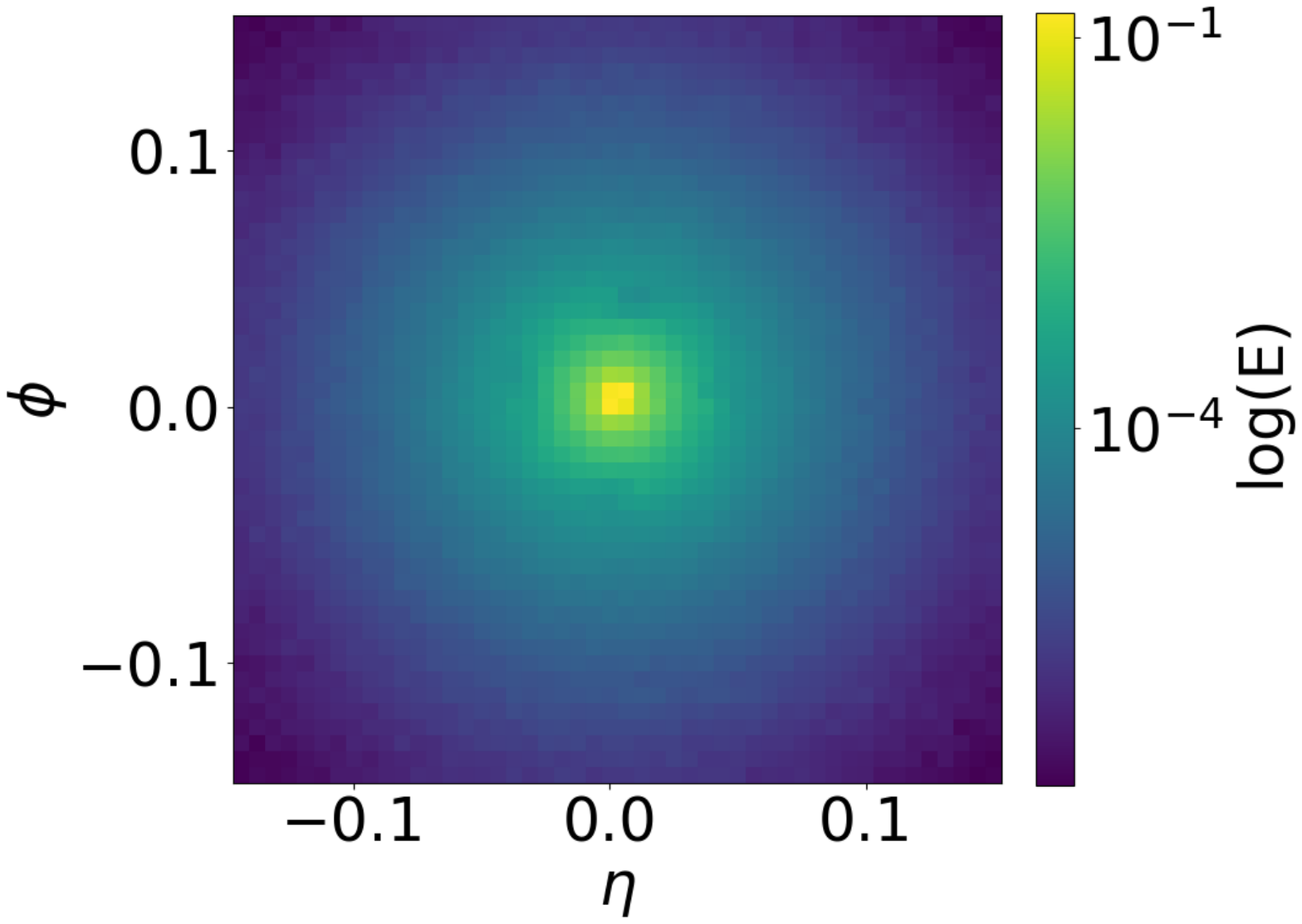}
        \caption{ Generated calorimeter deposition by SAVAE. }
        \label{fig:48x48_train}
    \end{subfigure}
    \hfill
    \begin{subfigure}[b]{0.45\textwidth}
        \centering
        \includegraphics[width=\textwidth]{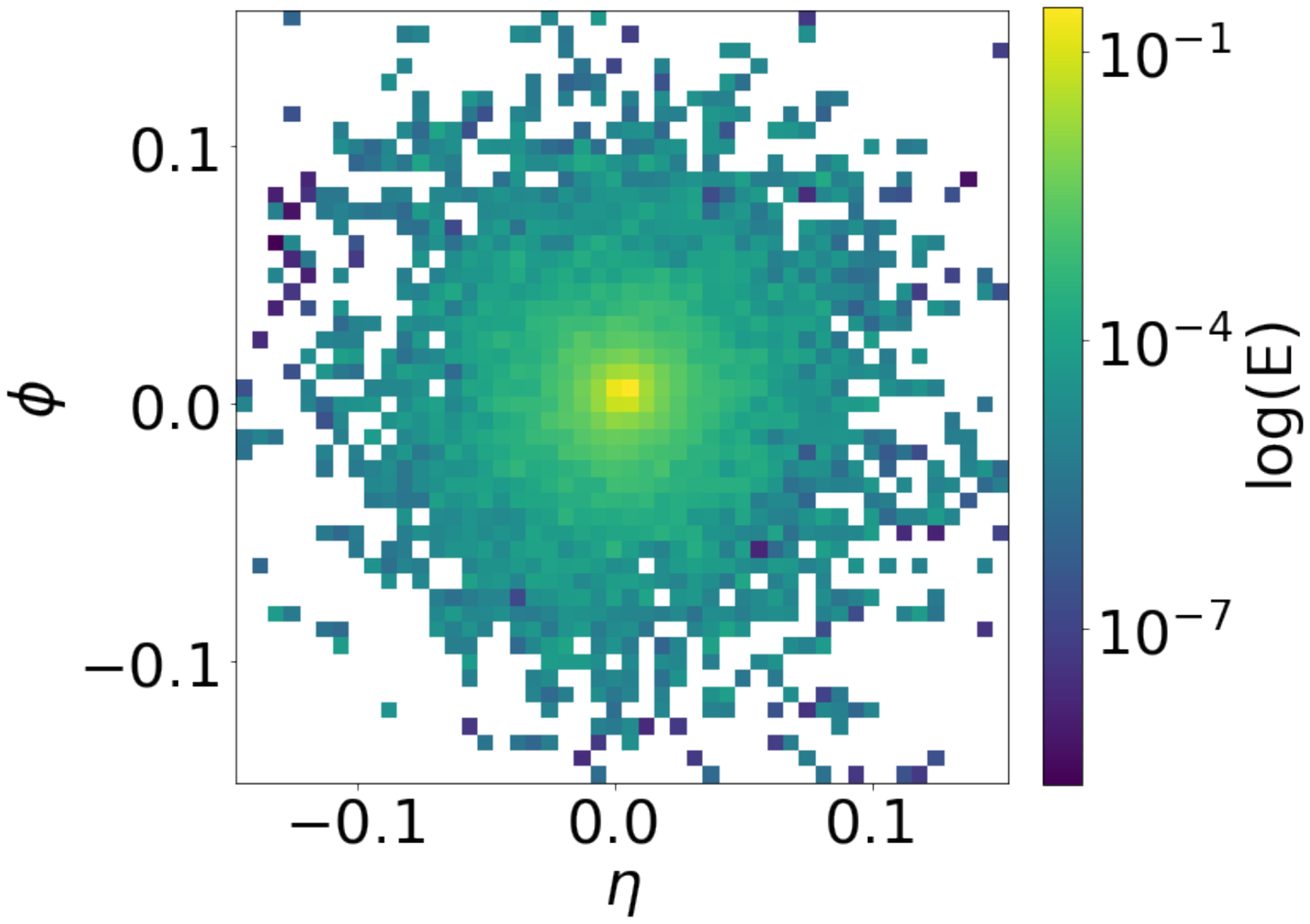}
        \caption{ Sample Geant4 training data.}
        \label{fig:48x48_gen}
    \end{subfigure}
    \caption{Sample generated and training data for the $(48,48)$ geometry. }
    \label{fig:48x48_imgs}
\end{figure}

\begin{figure}
    \centering
    \begin{subfigure}[b]{0.45\textwidth}
        \centering
        \includegraphics[width=\textwidth]{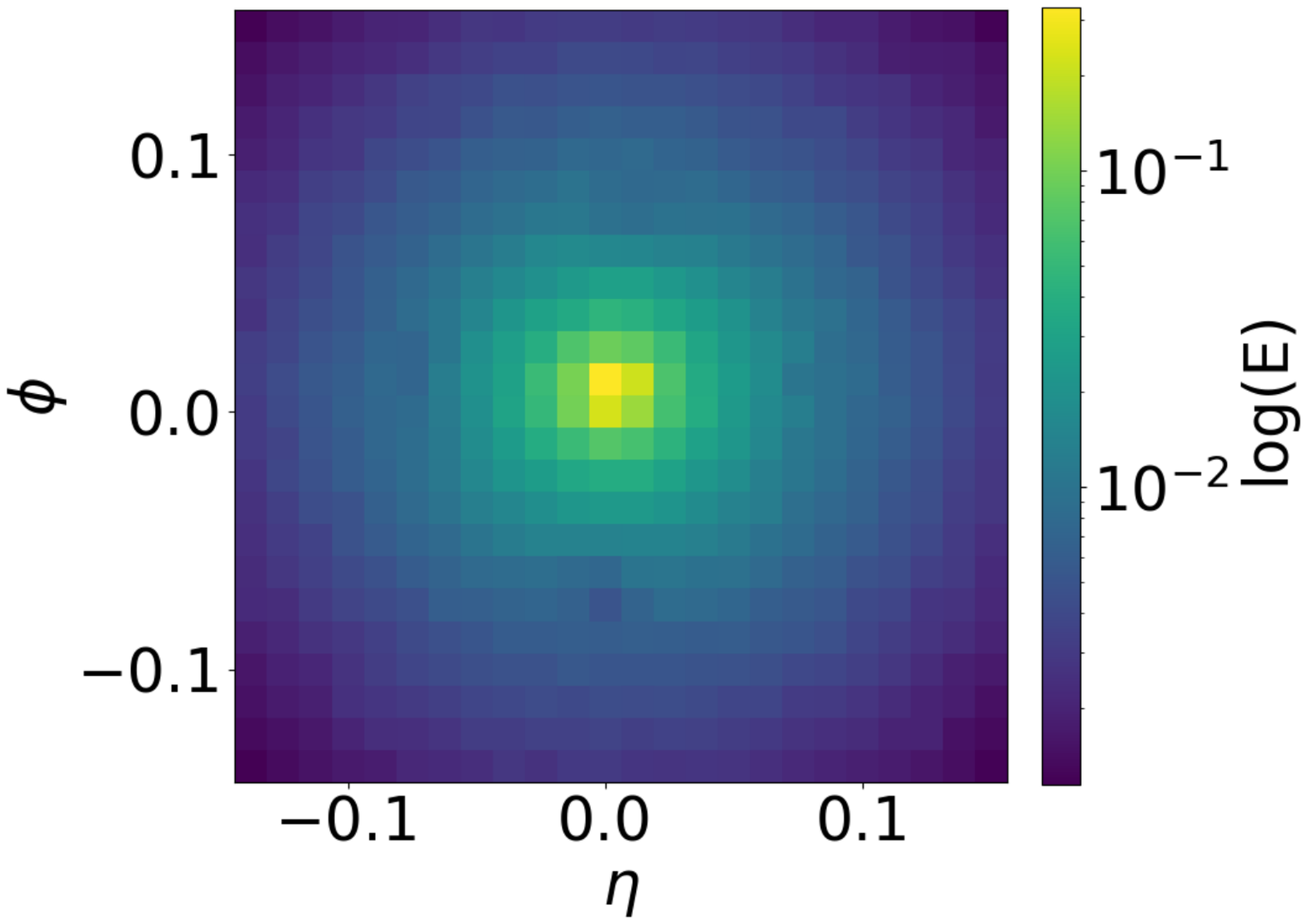}
        \caption{Generated calorimeter deposition by SAVAE.}
        \label{fig:24x24_interp}
    \end{subfigure}
    \hfill
    \begin{subfigure}[b]{0.45\textwidth}
        \centering
        \includegraphics[width=\textwidth]{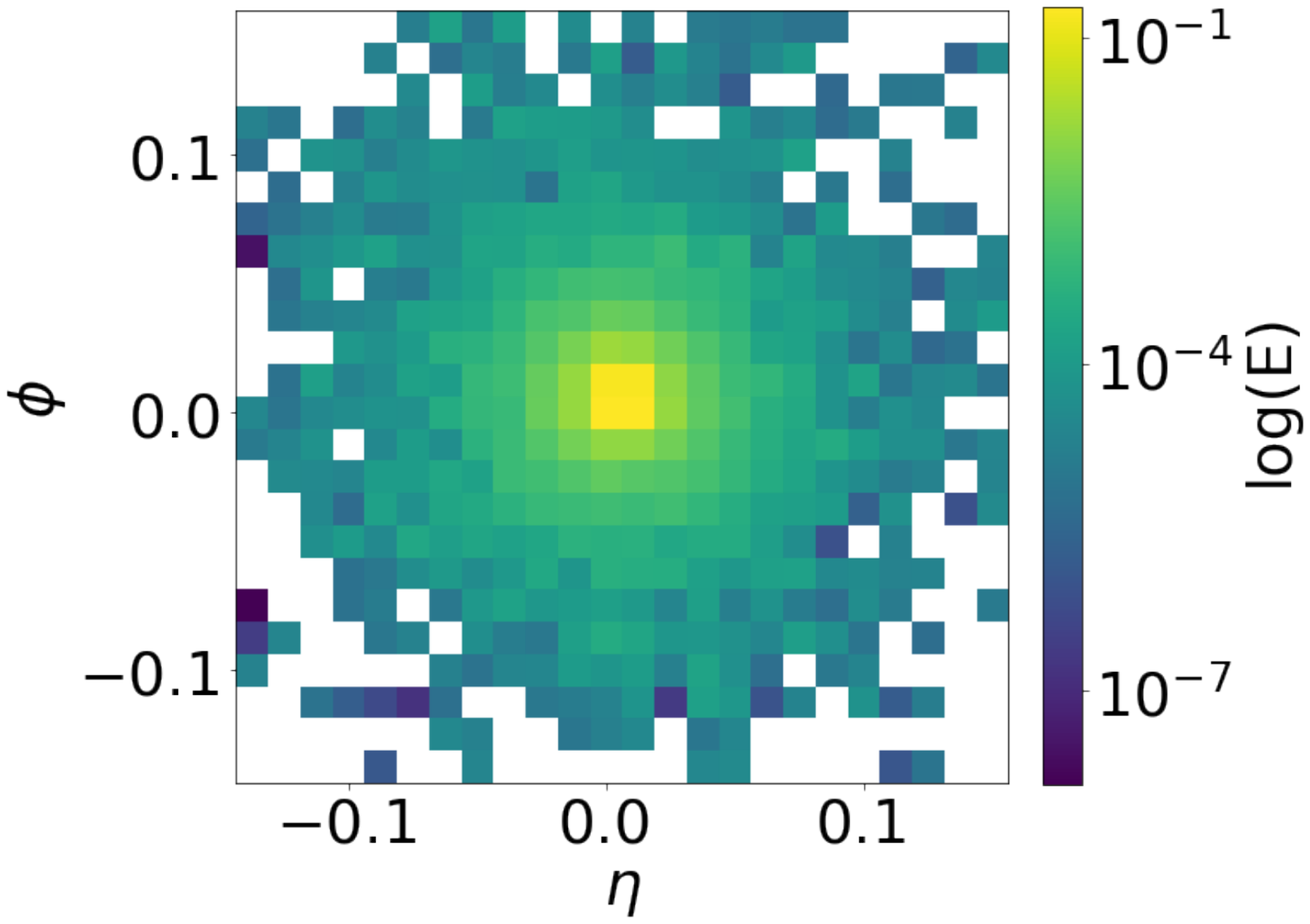}
        \caption{Sample Geant4 training data.}
        \label{fig:24x24_gen}
    \end{subfigure}
    \caption{Sample generated and training data for the $(24,24)$ geometry.}
    \label{fig:24x24_imgs}
\end{figure}

\begin{figure}
    \centering
    \begin{subfigure}[b]{0.45\textwidth}
        \centering
        \includegraphics[width=\textwidth]{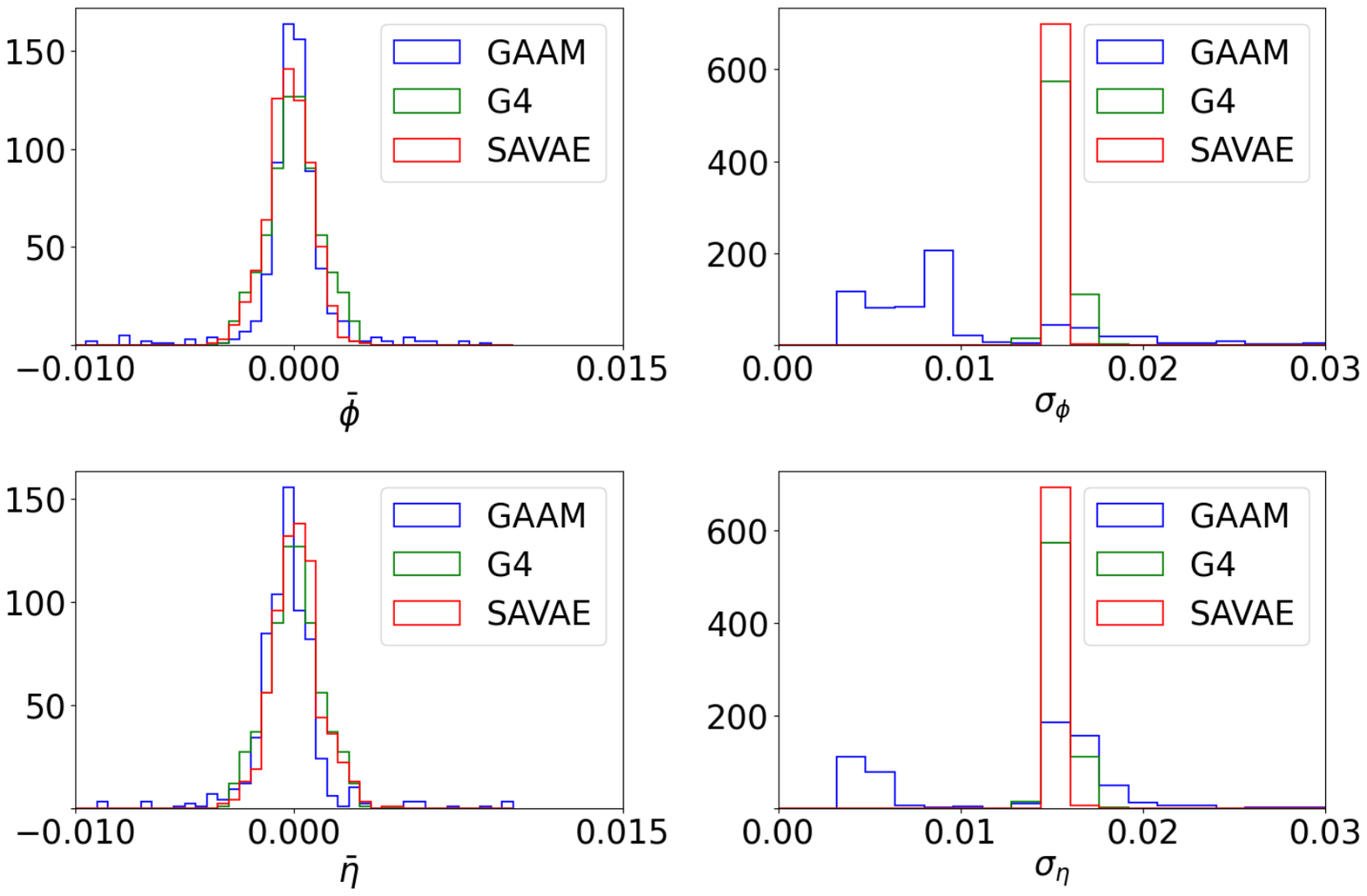}
        \caption{(48,48)}
        \label{fig:48x48_hists}
    \end{subfigure}
    \hfill
    \begin{subfigure}[b]{0.45\textwidth}
        \centering
        \includegraphics[width=\textwidth]{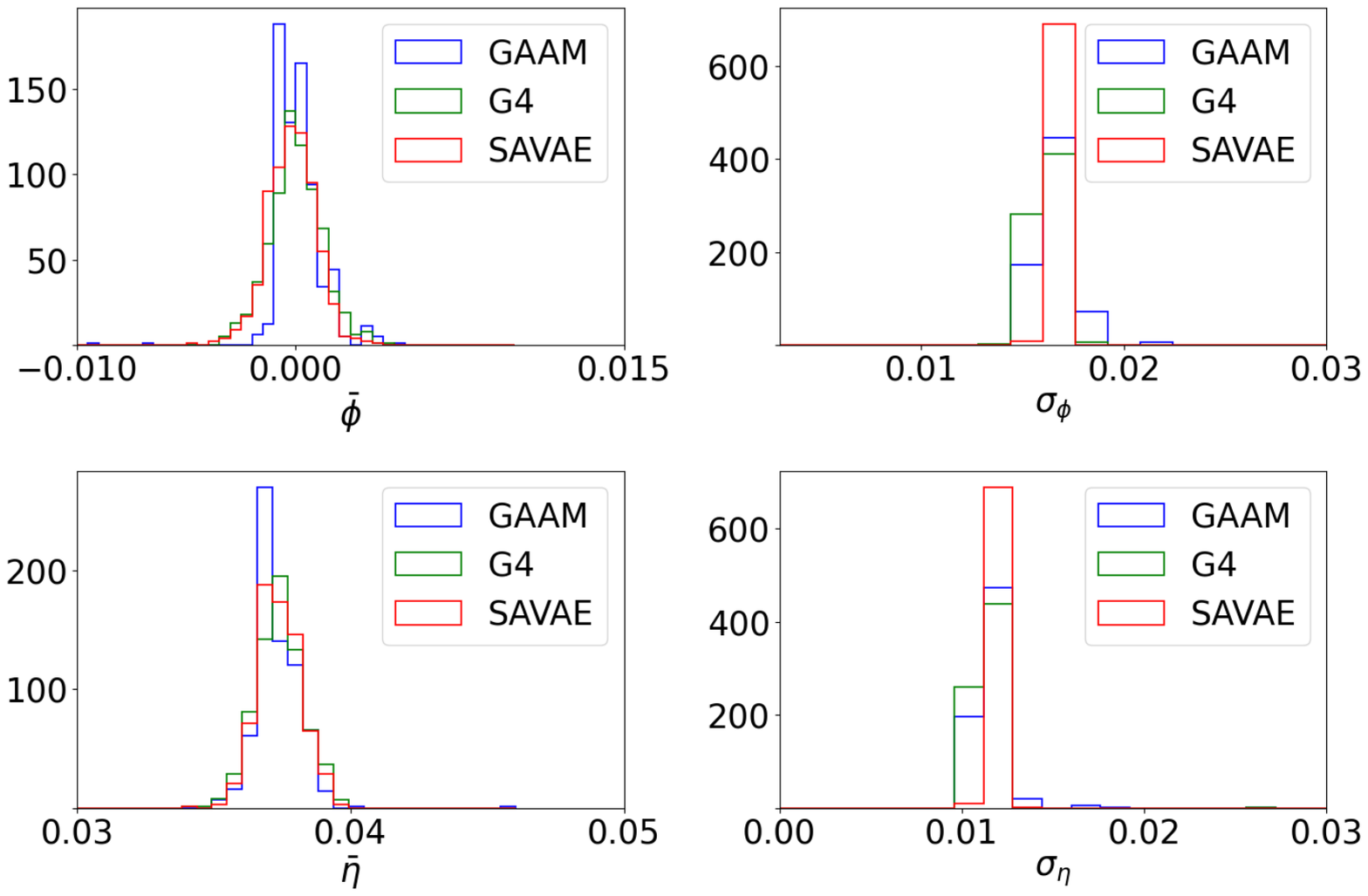}
        \caption{(36,48)}
        \label{fig:36x48_hists}
    \end{subfigure}
    \caption{Distributions of energy-weighted means, $\eta$ and $\phi$, and shower widths, $\sigma_{\eta}$ and $\sigma_{\phi}$, for SAVAE, GAAM, and Geant for samples generated in two training geometries, $(48,48)$ and $(36,48)$.}
    \label{fig:train_hists}
\end{figure}

\begin{figure}
    \centering
    \begin{subfigure}[b]{0.45\textwidth}
        \centering
        \includegraphics[width=\textwidth]{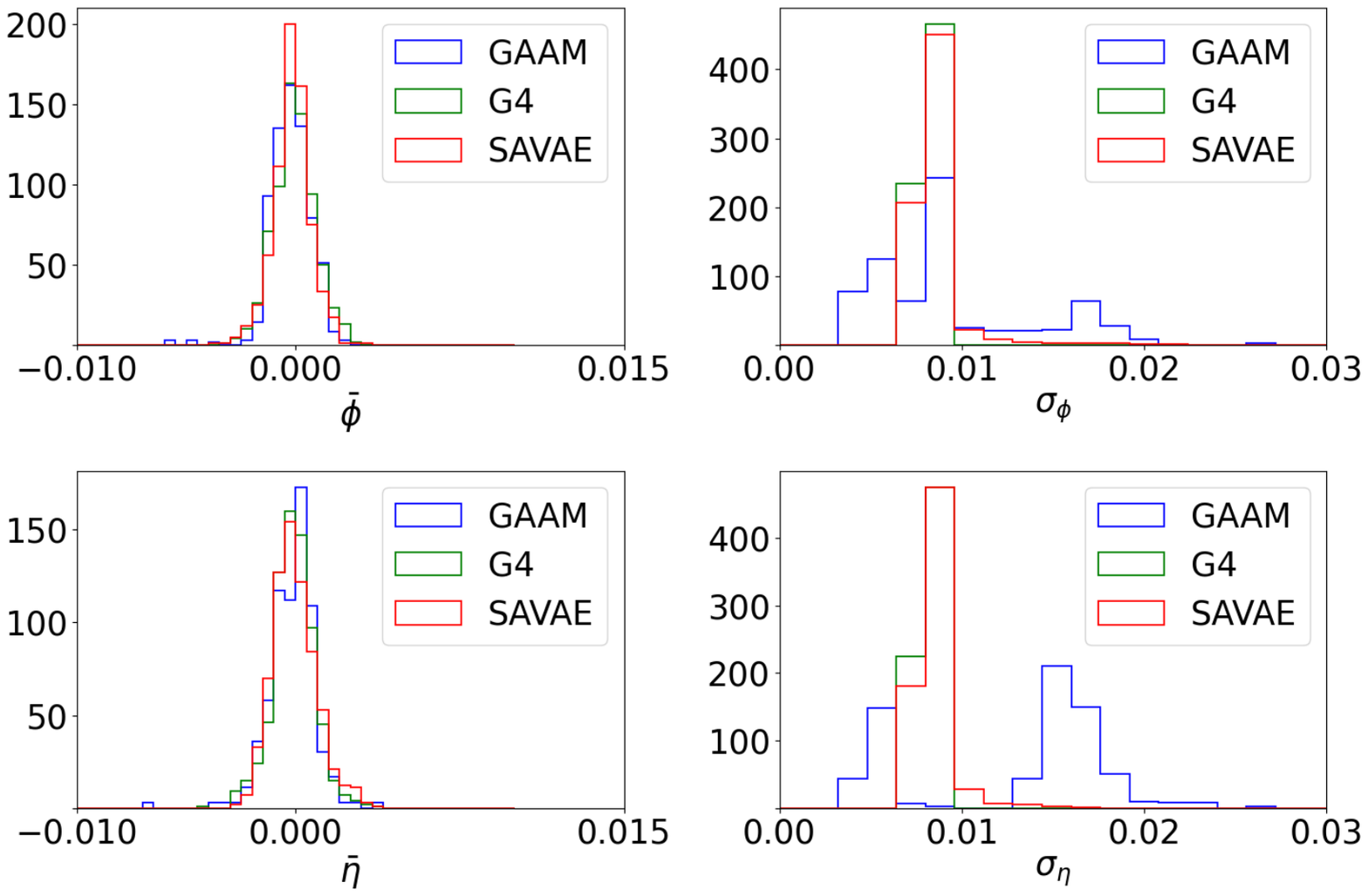}
        \caption{(24,24)}
        \label{fig:24x24_hists}
    \end{subfigure}
    \hfill
    \begin{subfigure}[b]{0.45\textwidth}
        \centering
        \includegraphics[width=\textwidth]{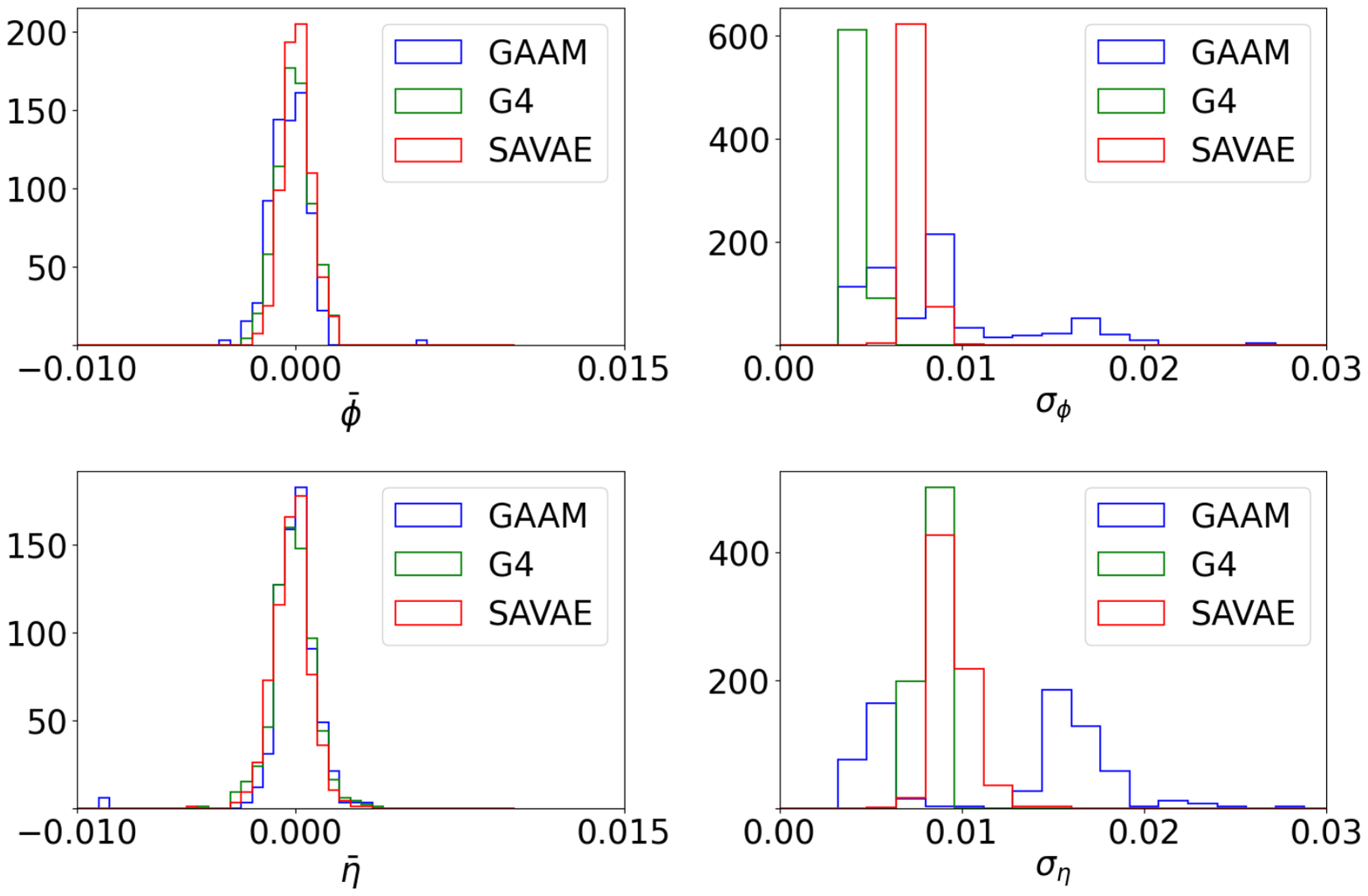}
        \caption{(24,12)}
        \label{fig:24x12_hists}
    \end{subfigure}
    \caption{Distributions of energy-weighted means, $\eta$ and $\phi$, and shower widths, $\sigma_{\eta}$ and $\sigma_{\phi}$, for SAVAE, GAAM, and Geant for samples generated in two interpolated geometries, $(24,24)$ and $(24,12)$.}
    \label{fig:interp_hists}
\end{figure}

\begin{table}
  \centering
   \caption{Wasserstein distances between the distributions of key physics quantities in the Geant4 training sample and in samples generated by GAAM or SAVAE.  Results are shown for four geometries, including two which are present in the training sample, $(48,48)$ and $(36,48)$, and two which are not seen during training, $(24,24)$ and $(24,12)$. The lowest value for each distribution comparison is bolded.}
  \label{wdist_train}
  \begin{tabular}{llccc}
    \toprule
    Segmentation   & Quantity   & \multicolumn{3}{c}{Wasserstein distance}\\
    & &     GAAM  & \ \ \ & SAVAE\\
    
    \midrule
    \multirow{4}{*}{(48,48)} & $\bar \eta$  & $5.95 \times 10^{-4}$  && $\bf{2.91 \times 10^{-4}}$ \\  
                                        & $\bar \phi$ & $4.05 \times 10^{-4}$  && $\bf{1.88 \times 10^{-4}}$  \\
                                        & $\sigma_{\eta}$ & $4.35 \times 10^{-3}$ && $\bf{4.56 \times 10^{-4}}$ \\
                                        & $\sigma_{\phi}$ & $7.07 \times 10^{-3}$ && $\bf{4.70 \times 10^{-4}}$ \\
    \midrule                                    
    \multirow{4}{*}{(36,48)} & $\bar \eta$  & $1.25 \times 10^{-2}$  && $\bf{1.58 \times 10^{-4}}$                                          \\  
                                        & $\bar \phi$ & $4.54 \times 10^{-4}$  && $\bf{8.37 \times 10^{-5}}$  \\
                                        & $\sigma_{\eta}$ & $1.18 \times 10^{-3}$ && $\bf{3.87 \times 10^{-4}}$  \\
                                        & $\sigma_{\phi}$ & $1.24 \times 10^{-3}$ && $\bf{2.60 \times 10^{-4}}$  \\
    \midrule
    \multirow{4}{*}{(24,24)} & $\bar \eta$  & $2.90 \times 10^{-4}$   && $\bf{7.47 \times 10^{-5}}$                                        \\
                                        & $\bar \phi$ & $6.21 \times 10^{-4}$    && $\bf{1.29 \times 10^{-4}}$ \\
                                        & $\sigma_{\eta}$ & $8.98 \times 10^{-3}$ && $\bf{4.07 \times 10^{-4}}$ \\
                                        & $\sigma_{\phi}$ & $4.22 \times 10^{-3}$ && $\bf{4.18 \times 10^{-4}}$ \\
    \midrule
    \multirow{4}{*}{(24,12)} & $\bar \eta$  & $4.77 \times 10^{-4}$  && $\bf{1.79 \times 10^{-4}}$ \\  
                                        & $\bar \phi$ & $1.02 \times 10^{-3}$ && $\bf{1.35 \times 10^{-4}}$\\
                                        & $\sigma_{\eta}$ & $8.98 \times 10^{-3}$ && $\bf{2.76 \times 10^{-3}}$ \\
                                        & $\sigma_{\phi}$ & $4.22 \times 10^{-3}$ && $\bf{1.46 \times 10^{-3}}$ \\
    \bottomrule
  \end{tabular}
\end{table}

Example calorimeter energy depositions generated by the SAVAE model are shown in Figs.~\ref{fig:48x48_imgs} and ~\ref{fig:24x24_imgs} for the $(48,48)$ and ($24,24$) geometries, respectively. Also shown are examples from the Geant4 training set.  Quantities of interest to physicists include the energy-weights means coordinate values, $\bar{\eta}\mathrm{\;and\;}\bar{\phi}$, and the shower width along each axis, $\sigma_{\eta}\mathrm{\;and\;}\sigma_{\phi}$, calculated as

\begin{equation} 
\label{eq_energy_weighted_mean}
\bar{\eta} = \frac{\sum_{i} \eta_{i}E_{i}}{\sum_{i} E_{i}},  \bar{\phi} = \frac{\sum_{i} \phi_{i}E_{i}}{\sum_{i} E_{i}} 
\end{equation}
where $E_{i}$ is the energy deposited in the $i^{th}$ cell. The shower widths, $\sigma_{\eta}$ and $\sigma_{\phi}$ are calculated as:
\begin{equation} 
\label{eq_shower_width}
\sigma_{\eta} = \sqrt{\frac{\sum_i E_{i} (\eta_i-\bar{\eta})^2}{\frac{(M-1)}{M} \sum_{i} E_{i}}}, \sigma_{\phi} = \sqrt{\frac{\sum_i E_{i} (\phi_i-\bar{\phi})^2}{\frac{(M-1)}{M} \sum_{i} E_{i}}}
\end{equation}
where $M$ is the number of cells with non-zero energy.
The energy-weighted means and shower widths are the most important properties of a shower to model correctly, as it has been shown that models that can correctly learn these properties also learn higher-level shower properties such as incident particle energy~\cite{liu2022geometryaware,Paganini_2018}.  Distributions of these quantities for geometries $(48,48)$ and $(36,48)$ are shown in Fig.~\ref{fig:train_hists}. Distributions for interpolation geometries $(24,24)$ and $(24,12)$, those not present in the training sample, are shown in Fig.~\ref{fig:interp_hists}.   Similarity between the distributions is measured via the Wasserstein distance (W-distance), as shown in Table~\ref{wdist_train}.  SAVAE performs better (a lower W-distance) for all training and interpolated geometries evaluated compared to GAAM, in some cases by two orders of magnitude. 

The SAVAE distributions for $\sigma$ are slightly narrower than those in Geant4, indicating less variation in energy deposition, especially in the training datasets.  The $\beta$-term can be decreased to encourage more variation in SAVAE $\sigma$-energy distribution; this specific model was optimized to reduce the W-distance for interpolated generation.

Generation time for SAVAE is much faster than for GAAM. Both GAAM and SAVAE samples were batch generated on GPU; generation takes 140.5 ms/shower for GAAM, compared to 0.568 ms/shower for SAVAE, as a result of the parallel nature of the self-attention architecture. GAAM was trained on two NVIDIA RTX 3090 GPUs, and SAVAE was trained on a single NVIDIA RTX A5000 GPU.

\section{Conclusion}
\label{conclusion}

This paper demonstrates the use of self-attention variable auto encoders (SAVAE) for fast simulation of variable geometry calorimeter segments. 

Compared to auto-regressive geometry-aware calorimeter simulation (GAAM), SAVAE offers faster training and sampling without the dependence on cell path ordering. SAVAE reliably simulates keystone physics distributions on  geometries unseen in training data, with smaller Wasserstein distances between generated and true distributions than achieved by GAAM. Compared to point-cloud calorimeter simulation, SAVAE offers a simplicity that reflects the regions of regularity of real calorimeters, which do not always require full flexibility for geometry specification.

Future work could include developing the capability to model multiple layers, and increasing the generality of the geometry specification, towards the goal of foundational ML models capable of adapting to diverse detectors, rather than training individual models for each calorimeter segmentation.


\section{Acknowledgements}

DW, AG and DS are supported by The Department of Energy Office of Science grant DE-SC0009920. DS is also supported by the HEPCAT Fellowship under contract DE-SC0022313. We are grateful to Benjamin Nachman for help with the Geant4 setup.

\bibliography{main}

\begin{thebibliography}{34}
\providecommand{\natexlab}[1]{#1}
\providecommand{\url}[1]{\texttt{#1}}
\expandafter\ifx\csname urlstyle\endcsname\relax
  \providecommand{\doi}[1]{doi: #1}\else
  \providecommand{\doi}{doi: \begingroup \urlstyle{rm}\Url}\fi

\bibitem[Cranmer et~al.(2020)Cranmer, Brehmer, and Louppe]{Cranmer:2019eaq}
Kyle Cranmer, Johann Brehmer, and Gilles Louppe.
\newblock {The frontier of simulation-based inference}.
\newblock \emph{Proc. Nat. Acad. Sci.}, 117\penalty0 (48):\penalty0 30055--30062, 2020.
\newblock \doi{10.1073/pnas.1912789117}.

\bibitem[Agostinelli et~al.(2003)]{GEANT4:2002zbu}
S.~Agostinelli et~al.
\newblock {GEANT4--a simulation toolkit}.
\newblock \emph{Nucl. Instrum. Meth. A}, 506:\penalty0 250--303, 2003.
\newblock \doi{10.1016/S0168-9002(03)01368-8}.

\bibitem[Allison et~al.(2016)]{Allison:2016lfl}
J.~Allison et~al.
\newblock {Recent developments in Geant4}.
\newblock \emph{Nucl. Instrum. Meth. A}, 835:\penalty0 186--225, 2016.
\newblock \doi{10.1016/j.nima.2016.06.125}.

\bibitem[Allison et~al.(2006)]{Allison:2006ve}
John Allison et~al.
\newblock {Geant4 developments and applications}.
\newblock \emph{IEEE Trans. Nucl. Sci.}, 53:\penalty0 270, 2006.
\newblock \doi{10.1109/TNS.2006.869826}.

\bibitem[Baldi et~al.(2014)Baldi, Sadowski, and Whiteson]{baldi2014searching}
Pierre Baldi, Peter Sadowski, and Daniel Whiteson.
\newblock Searching for exotic particles in high-energy physics with deep learning.
\newblock \emph{Nature communications}, 5\penalty0 (1):\penalty0 4308, 2014.

\bibitem[Baldi(2021)]{baldi2021deep}
Pierre Baldi.
\newblock \emph{Deep learning in science}.
\newblock Cambridge University Press, 2021.

\bibitem[Calafiura et~al.(2022)Calafiura, Rousseau, and Terao]{calafiura2022artificial}
Paolo Calafiura, David Rousseau, and Kazuhiro Terao.
\newblock \emph{Artificial Intelligence for High Energy Physics}.
\newblock World Scientific, 2022.

\bibitem[Lu et~al.(2022)Lu, Romero, Fenton, Whiteson, and Baldi]{lu2022resolving}
Yadong Lu, Alexis Romero, Michael~James Fenton, Daniel Whiteson, and Pierre Baldi.
\newblock Resolving extreme jet substructure.
\newblock \emph{Journal of High Energy Physics}, 2022\penalty0 (8):\penalty0 1--30, 2022.

\bibitem[Shmakov et~al.(2023)Shmakov, Greif, Fenton, Ghosh, Baldi, and Whiteson]{shmakov2023end}
Alexander Shmakov, Kevin Greif, Michael Fenton, Aishik Ghosh, Pierre Baldi, and Daniel Whiteson.
\newblock End-to-end latent variational diffusion models for inverse problems in high energy physics.
\newblock \emph{arXiv preprint arXiv:2305.10399}, 2023.

\bibitem[Paganini et~al.(2018{\natexlab{a}})Paganini, de~Oliveira, and Nachman]{Paganini:2017dwg}
Michela Paganini, Luke de~Oliveira, and Benjamin Nachman.
\newblock {CaloGAN : Simulating 3D high energy particle showers in multilayer electromagnetic calorimeters with generative adversarial networks}.
\newblock \emph{Phys. Rev.}, D97\penalty0 (1):\penalty0 014021, 2018{\natexlab{a}}.
\newblock \doi{10.1103/PhysRevD.97.014021}.

\bibitem[{Vallecorsa, Sofia} et~al.(2019){Vallecorsa, Sofia}, {Carminati, Federico}, and {Khattak, Gulrukh}]{3dgan_epj}
{Vallecorsa, Sofia}, {Carminati, Federico}, and {Khattak, Gulrukh}.
\newblock 3d convolutional gan for fast simulation.
\newblock \emph{EPJ Web Conf.}, 214:\penalty0 02010, 2019.
\newblock \doi{10.1051/epjconf/201921402010}.
\newblock URL \url{https://doi.org/10.1051/epjconf/201921402010}.

\bibitem[Buhmann et~al.(2021)Buhmann, Diefenbacher, Eren, Gaede, Kasieczka, Korol, and Kr\"uger]{Buhmann:2020pmy}
Erik Buhmann, Sascha Diefenbacher, Engin Eren, Frank Gaede, Gregor Kasieczka, Anatolii Korol, and Katja Kr\"uger.
\newblock {Getting High: High Fidelity Simulation of High Granularity Calorimeters with High Speed}.
\newblock \emph{Comput. Softw. Big Sci.}, 5\penalty0 (1):\penalty0 13, 2021.
\newblock \doi{10.1007/s41781-021-00056-0}.

\bibitem[Krause and Shih(2021{\natexlab{a}})]{Krause:2021ilc}
Claudius Krause and David Shih.
\newblock {CaloFlow: Fast and Accurate Generation of Calorimeter Showers with Normalizing Flows}, 6 2021{\natexlab{a}}.

\bibitem[Krause and Shih(2021{\natexlab{b}})]{Krause:2021wez}
Claudius Krause and David Shih.
\newblock {CaloFlow II: Even Faster and Still Accurate Generation of Calorimeter Showers with Normalizing Flows}, 10 2021{\natexlab{b}}.

\bibitem[Mikuni and Nachman(2022)]{Mikuni:2022xry}
Vinicius Mikuni and Benjamin Nachman.
\newblock {Score-based Generative Models for Calorimeter Shower Simulation}, 6 2022.

\bibitem[Amram and Pedro(2023)]{amram2023denoisingdiffusionmodelsgeometry}
Oz~Amram and Kevin Pedro.
\newblock Denoising diffusion models with geometry adaptation for high fidelity calorimeter simulation, 2023.
\newblock URL \url{https://arxiv.org/abs/2308.03876}.

\bibitem[Aad et~al.(2022)]{ATLAS:2021pzo}
Georges Aad et~al.
\newblock {AtlFast3: the next generation of fast simulation in ATLAS}.
\newblock \emph{Comput. Softw. Big Sci.}, 6:\penalty0 7, 2022.
\newblock \doi{10.1007/s41781-021-00079-7}.

\bibitem[ATL(2022)]{ATLAS:2022jhk}
{Deep generative models for fast photon shower simulation in ATLAS}, 10 2022.

\bibitem[Erdmann et~al.(2019)Erdmann, Glombitza, and Quast]{Erdmann:2018jxd}
Martin Erdmann, Jonas Glombitza, and Thorben Quast.
\newblock {Precise simulation of electromagnetic calorimeter showers using a Wasserstein Generative Adversarial Network}.
\newblock \emph{Comput. Softw. Big Sci.}, 3\penalty0 (1):\penalty0 4, 2019.
\newblock \doi{10.1007/s41781-018-0019-7}.

\bibitem[Ratnikov(2020)]{Ratnikov:2020dcm}
Fedor Ratnikov.
\newblock {Generative Adversarial Networks for LHCb Fast Simulation}.
\newblock \emph{EPJ Web Conf.}, 245:\penalty0 02026, 2020.
\newblock \doi{10.1051/epjconf/202024502026}.

\bibitem[Hashemi et~al.(2023)Hashemi, Hartmann, Sharifzadeh, Kahn, and Kuhr]{Hashemi:2023ruu}
Hosein Hashemi, Nikolai Hartmann, Sahand Sharifzadeh, James Kahn, and Thomas Kuhr.
\newblock {Ultra-High-Resolution Detector Simulation with Intra-Event Aware GAN and Self-Supervised Relational Reasoning}, 3 2023.

\bibitem[Liu et~al.(2022)Liu, Ghosh, Smith, Baldi, and Whiteson]{liu2022geometryaware}
Junze Liu, Aishik Ghosh, Dylan Smith, Pierre Baldi, and Daniel Whiteson.
\newblock Geometry-aware autoregressive models for calorimeter shower simulations.
\newblock \emph{JINST}, 2022.

\bibitem["Aad and others"(2022)]{dg2024}
G.~"Aad and others".
\newblock Deep generative models for fast photon shower simulation in atlas.
\newblock \emph{Computing and Software for Big Science}, 8\penalty0 (1), 2022.
\newblock ISSN 2510-2044.
\newblock \doi{10.1007/s41781-023-00106-9}.
\newblock URL \url{http://dx.doi.org/10.1007/s41781-023-00106-9}.

\bibitem[Baldi et~al.(2016)Baldi, Cranmer, Faucett, Sadowski, and Whiteson]{Baldi_2016}
Pierre Baldi, Kyle Cranmer, Taylor Faucett, Peter Sadowski, and Daniel Whiteson.
\newblock Parameterized neural networks for high-energy physics.
\newblock \emph{The European Physical Journal C}, 76\penalty0 (5), April 2016.
\newblock ISSN 1434-6052.
\newblock \doi{10.1140/epjc/s10052-016-4099-4}.
\newblock URL \url{http://dx.doi.org/10.1140/epjc/s10052-016-4099-4}.

\bibitem[Kansal et~al.(2021)Kansal, Duarte, Su, Orzari, Tomei, Pierini, Touranakou, Vlimant, and Gunopulos]{Kansal:2021cqp}
Raghav Kansal, Javier Duarte, Hao Su, Breno Orzari, Thiago Tomei, Maurizio Pierini, Mary Touranakou, Jean-Roch Vlimant, and Dimitrios Gunopulos.
\newblock {Particle Cloud Generation with Message Passing Generative Adversarial Networks}.
\newblock In \emph{{35th Conference on Neural Information Processing Systems}}, 6 2021.

\bibitem[Mikuni et~al.(2023)Mikuni, Nachman, and Pettee]{Mikuni:2023dvk}
Vinicius Mikuni, Benjamin Nachman, and Mariel Pettee.
\newblock {Fast Point Cloud Generation with Diffusion Models in High Energy Physics}.
\newblock \emph{American Physical Society ({APS})}, 4 2023.

\bibitem[Leigh et~al.(2023)Leigh, Sengupta, Qu\'etant, Raine, Zoch, and Golling]{Leigh:2023toe}
Matthew Leigh, Debajyoti Sengupta, Guillaume Qu\'etant, John~Andrew Raine, Knut Zoch, and Tobias Golling.
\newblock {PC-JeDi: Diffusion for Particle Cloud Generation in High Energy Physics}, 3 2023.

\bibitem[Paganini et~al.(2018{\natexlab{b}})Paganini, de~Oliveira, and Nachman]{Paganini_2018}
Michela Paganini, Luke de~Oliveira, and Benjamin Nachman.
\newblock {CaloGAN}: Simulating 3d high energy particle showers in multilayer electromagnetic calorimeters with generative adversarial networks.
\newblock \emph{Physical Review D}, 97\penalty0 (1), 2018{\natexlab{b}}.
\newblock \doi{10.1103/PhysRevD.97.014021}.

\bibitem[Scarselli et~al.(2009)Scarselli, Gori, Tsoi, Hagenbuchner, and Monfardini]{4700287}
Franco Scarselli, Marco Gori, Ah~Chung Tsoi, Markus Hagenbuchner, and Gabriele Monfardini.
\newblock The graph neural network model, 2009.

\bibitem[Kingma and Welling(2022)]{kingma2022autoencoding}
Diederik~P Kingma and Max Welling.
\newblock Auto-encoding variational bayes, 2022.

\bibitem[Salimans et~al.(2016)Salimans, Goodfellow, Zaremba, Cheung, Radford, and Chen]{salimans2016improved}
Tim Salimans, Ian Goodfellow, Wojciech Zaremba, Vicki Cheung, Alec Radford, and Xi~Chen.
\newblock Improved techniques for training gans, 2016.

\bibitem[Higgins et~al.(2017)Higgins, Matthey, Pal, Burgess, Glorot, Botvinick, Mohamed, and Lerchner]{higgins2017betavae}
Irina Higgins, Loic Matthey, Arka Pal, Christopher Burgess, Xavier Glorot, Matthew Botvinick, Shakir Mohamed, and Alexander Lerchner.
\newblock beta-{VAE}: Learning basic visual concepts with a constrained variational framework.
\newblock In \emph{International Conference on Learning Representations}, 2017.
\newblock URL \url{https://openreview.net/forum?id=Sy2fzU9gl}.

\bibitem[Lu et~al.(2021)Lu, Collado, Whiteson, and Baldi]{lu2021sparse}
Yadong Lu, Julian Collado, Daniel Whiteson, and Pierre Baldi.
\newblock Sparse autoregressive models for scalable generation of sparse images in particle physics.
\newblock \emph{Physical Review D}, 103\penalty0 (3):\penalty0 036012, 2021.

\bibitem[Zhang et~al.(2019)Zhang, Goodfellow, Metaxas, and Odena]{zhang2019selfattention}
Han Zhang, Ian Goodfellow, Dimitris Metaxas, and Augustus Odena.
\newblock Self-attention generative adversarial networks, 2019.

\end{thebibliography}

\appendix
\section{Geometry-Specific Training}

\begin{table}
  \centering
   \caption{Wasserstein distances between the distributions of key physics quantities in the Geant4 training sample and in samples generated by SAVAE or the geometry-specific baseline model.  Results are shown for four geometries, including two which are present in the training sample, $(48,48)$ and $(36,48)$, and two which are not seen during training, $(24,24)$ and $(24,12)$. The lowest value for each distribution comparison is bolded.}
  \label{wdist_specific}
  \begin{tabular}{llccc}
    \toprule
    Segmentation   & Quantity   & \multicolumn{3}{c}{Wasserstein distance}\\
    & &     SAVAE  & \ \ \ & Geometry-Specific \\
    
    \midrule
    \multirow{4}{*}{(48,48)} & $\bar \eta$  & $2.91 \times 10^{-4}$ && $\bf{1.82 \times 10^{-4}}$ \\  
                                        & $\bar \phi$ & $1.88 \times 10^{-4}$ && $\bf{1.06 \times 10^{-4}}$  \\
                                        & $\sigma_{\eta}$ & $4.56 \times 10^{-4}$ && $\bf{1.79 \times 10^{-4}}$\\
                                        & $\sigma_{\phi}$ & $4.70 \times 10^{-4}$ && $\bf{4.03 \times 10^{-4}}$\\
    \midrule                                    
    \multirow{4}{*}{(36,48)} & $\bar \eta$  & $1.58 \times 10^{-4}$ && $\bf{1.37 \times 10^{-4}}$\\  
                                        & $\bar \phi$ & $\bf{8.37 \times 10^{-5}}$ && $0.48 \times 10^{-4}$ \\
                                        & $\sigma_{\eta}$ & $3.87 \times 10^{-4}$ && $\bf{1.01 \times 10^{-4}}$ \\
                                        & $\sigma_{\phi}$ & $2.60 \times 10^{-4}$ && $\bf{2.48 \times 10^{-4}}$ \\
    \midrule
    \multirow{4}{*}{(24,24)} & $\bar \eta$  & $\bf{7.47 \times 10^{-5}}$  && $7.89 \times 10^{-4}$ \\
                                        & $\bar \phi$ & $1.29 \times 10^{-4}$ && $\bf{1.12 \times 10^{-4}}$\\
                                        & $\sigma_{\eta}$ &  $4.07 \times 10^{-4}$ && $\bf{3.98 \times 10^{-4}}$\\
                                        & $\sigma_{\phi}$ & $4.18 \times 10^{-4}$ && $\bf{3.77 \times 10^{-4}}$\\
    \midrule
    \multirow{4}{*}{(24,12)} & $\bar \eta$  &  $1.79 \times 10^{-4}$ && $\bf{4.23 \times 10^{-4}}$\\  
                                        & $\bar \phi$ & $1.35 \times 10^{-4}$ && $\bf{0.99 \times 10^{-5}}$\\
                                        & $\sigma_{\eta}$ & $2.76 \times 10^{-3}$ && $\bf{1.64 \times 10^{-3}}$ \\
                                        & $\sigma_{\phi}$ & $1.46 \times 10^{-3}$ && $\bf{1.39 \times 10^{-3}}$\\
    \bottomrule
  \end{tabular}
\end{table}

Alongside SAVAE and GAAM results, W-distances are calculated between Geant4 distributions and geometry-specific models. These models are identical in architecture to SAVAE, but are trained only on individual geometries on which they are tested. Because each of these models only need to learn one geometry at a time, their W-distances serve as an idealized benchmark to assess if SAVAE loses accuracy from handling multiple geometries. While the SAVAE model generally yields slightly higher W-distances, it performs comparably to the baseline and even matches its accuracy in several metrics. The comparison between SAVAE and each geometry specific model is shown in Table~\ref{wdist_specific}.

\end{document}